# SCALING OF DISCRETE ELEMENT MODEL PARAMETERS FOR COHESIONLESS AND COHESIVE SOLID

Subhash C Thakur[+], Jin Y. Ooi[*] and Hossein Ahmadian[+]

[+] Newcastle Innovation Centre, Procter and Gamble Technical Centre Ltd, Newcastle upon Tyne, NE12 9BZ, UK.

[*] School of Engineering, University of Edinburgh, Kings Buildings, Edinburgh EH9 3JL, UK

## Abstract

One of the major shortcomings of discrete element modelling (DEM) is the computational cost required when the number of particles is huge, especially for fine powders and/or industry scale simulations. This study investigates the scaling of model parameters that is necessary to produce scale independent predictions for cohesionless and cohesive solid under quasi-static simulation of confined compression and unconfined compression to failure. A bilinear elasto-plastic adhesive frictional contact model was used. The results show that contact stiffness (both normal and tangential) for loading and unloading scales linearly with the particle size and the adhesive force scales very well with the square of the particle size. This scaling law would allow scaled up particle DEM model to exhibit bulk mechanical loading response in uniaxial test that is similar to a material comprised of much smaller particles. This is a first step towards a mesoscopic representation of a cohesive powder that is phenomenological based to produce the key bulk characteristics of a cohesive solid and has the potential to gain considerable computational advantage for industry scale DEM simulations.

**1. Introduction**

The discrete element modelling originally developed by Cundall and Strack (1979) has increasingly been used to model many problems involving discrete phenomena including powder packing (Parteli et al., 2014; Yen and Chaki, 1992), compaction (Luding, 2008; Sheng et al., 2004a) powder flow (Moreno-atanasio et al., 2005;Singh et al., 2014;Thakur et al., 2014; Thakur et al., 2014), rotating drum (Walton and Johnson, 2010), mixing (Chaudhuri et al., 2006), hopper flow (González-Montellano et al., 2011; Ketterhagen et al., 2009), fluidized bed (Xu, 1997), pneumatic conveying (Ebrahimi et al., 2014; Sakai and Koshizuka, 2009) and many others. A detailed report on the applications of DEM can be found in the review paper by Zhu et al. (2008). The DEM simulations of the aforementioned phenomena have given many significant insights into the microscopic details at particle level and useful information to understand complex behaviour exhibited by granular material. For fine particles, one major shortcoming of DEM simulations for practical applications is the challenge of modelling very small particles. Even the smallest industrial processes involve interaction of trillions of fine particles, and it becomes computationally impossible and impractical to account for every individual realistically sized particles.

There are several possible solutions (Mio et al., 2009) for the speed-up of DEM simulation, such as optimization of the hardware and the software, including improved DEM algorithm, parallel computing, and simplifying the calculation process. Common ways to simplify the calculation process are done, for example, using a lower spring stiffness, using mono-sized particles, using a cut-off distance for long range forces (Mio et al., 2009) etc. Other possibilities are the use of higher particle density in quasi-static simulation (Sheng et al., 2004b) known as density scaling, reduction of number of particles by scaling the system size down or scaling up the size of parti-cle. Poschel et al. (2001) proposed a general approach to scale down the experiments to laboratory size. They found that the dynamics of their granular system changed if all sizes were scaled by a constant factor, but leaving the material properties the same.

Poschel's approach is more suitable for problems where an original physical problem is scaled down to a laboratory model in an attempt to obtain a physical model of the problem. This approach may not reduce the computational time in DEM modeling because the number of particles still remains the same and the particle size is also reduced. Another scaling approach is to use larger size elements (particles) to reduce the number of particles whilst keeping the original system size the same. One possible solution is to use larger size elements (particles) to reduce the number of particles whilst keeping the original system size the same, however, this would violate geometric similarity and may introduce some error in the bulk response as reported in Feng et al.(2007). The major issue in this kind of approach is to adjust DEM model parameters such that large particle DEM simulation result exhibits the same dynamic and static properties as small size realistic particles. This approach is sometimes referred to as coarse graining approach and has been used by a few researchers in the field of cavity filling (Bierwisch et al., 2009), pneumatic conveying (Sakai and Koshizuka, 2009), and rotary drum (Walton and Johnson, 2010).

This study investigates the scaling of model parameters that is necessary to produce scale independent predictions for cohesionless and cohesive powder under quasi-static 3D simulation of confined compression and unconfined compression to failure. The target is to develop DEM model with scaled up DEM particle to exhibit the compression and shearing bulk behaviour in a uniaxial test exhibited by powders.

## 2. DEM model and theoretical background for scaling

### 2.1 DEM contact model

A DEM contact model based on the physical phenomena observed in adhesive contact experiments has been proposed (Jones, 2003). When two particles or agglomerates are pressed together, they undergo elastic and plastic deformations and the pull-off (adhesive) force increases with an increase of the plastic contact area. Figure 1 shows the contact model in its full generic form which captures the key elements of the frictional-adhesive contact mechanics in that: $f_0$ provides the van der Waals type pull-off forces; $k_1$ and $k_2$ provides the elasto-plastic contact; $k_{adh}$ provides the load dependent adhesion; the exponent $n$ provides the nonlinearity and the resulting contact plasticity defines the total contact adhesion. The model is thus expected to be capable of modelling fine powders to study phenomena such as agglomeration, attrition and flow.

The schematic diagram of normal force-overlap ($f_n - \delta$) for this model is shown in Figure 2. When n=1 the model becomes linear (Figure1b) and similar to existing contact models (Luding, 2008; Singh, A,Magnanimo, V., and Luding, 2015; Walton and Johnson, 2009). The linear version of the contact model is used in this study. The details of the contact model are presented elsewhere (Thakur et al., 2014).

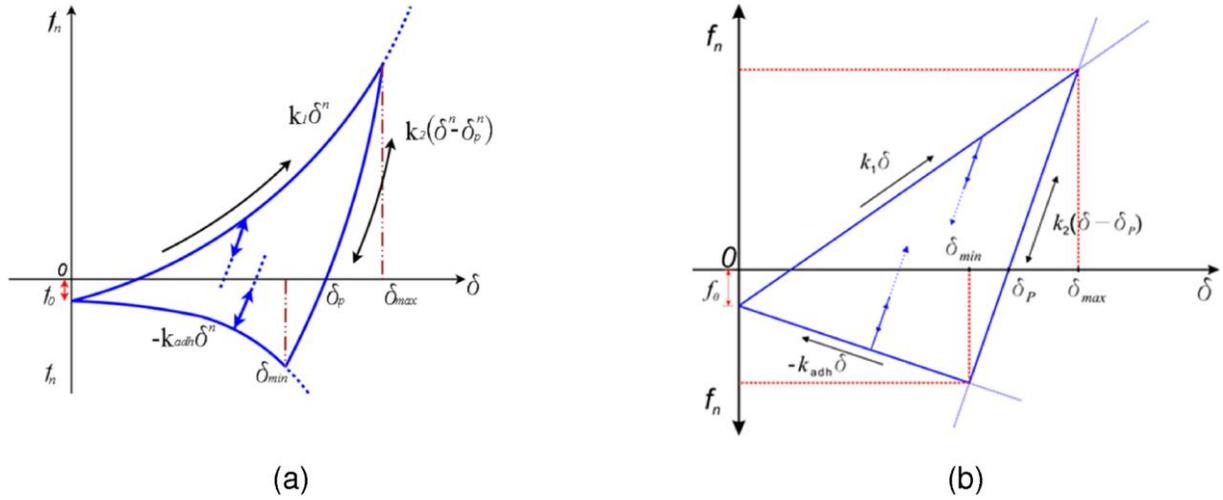

**Figure 1 Normal contact force-displacement function for the implemented model.**

This contact model has been implemented through the API in EDEM® v2.3, a commercial DEM code developed by DEM Solutions Ltd (2010). The total contact normal force, $f_n$, is the sum of the hysteretic spring force, $f_{hys}$, and the normal damping force, $f_{nd}$:

$$\boldsymbol{f_n} = (f_{hys} + f_{nd})\boldsymbol{u}, \tag{1}$$

where, $\boldsymbol{u}$ is the unit normal vector pointing from the contact point to the particle centre. The force-overlap relationship for normal contact, $f_{hys}$, is mathematically expressed by equation 2.

$$f_{hys} = \begin{cases} f_0 + k_1 \delta^n & \text{if } k_2(\delta^n - \delta_p^n) \geq k_1 \delta^n \\ f_0 + k_2(\delta^n - \delta_p^n) & \text{if } k_1 \delta^n > k_2(\delta^n - \delta_p^n) > -k_{adh}\delta^n \\ f_0 - k_{adh}\delta^n & \text{if } -k_{adh}\delta^n \geq k_2(\delta^n - \delta_p^n) \end{cases} \tag{2}$$

The normal damping force, $f_{nd}$, is given by:

$$f_{nd} = \beta_n v_n \qquad (3)$$

where $v_n$ is the magnitude of the relative normal velocity, and $\beta_n$ is the normal dashpot coefficient expressed as:

$$\beta_n = \sqrt{\frac{4m^* k_1}{1+\left(\dfrac{\pi}{\ln e}\right)^2}} \qquad (4)$$

with the equivalent mass of the particles $m^*$ defined as $(m_i m_j / m_i+m_j)$, where $m$ is the mass of the respective particles, and the coefficient of restitution $e$ defined in the simulation.

The contact tangential force, $f_t$, is given by the sum of tangential spring force, $f_{ts}$, and tangential damping force, $f_{td}$, as given by:

$$f_t = (f_{ts} + f_{td}). \qquad (5)$$

The tangential spring force is expressed in incremental terms:

$$f_{ts} = f_{ts(n-1)} + \Delta f_{ts}, \qquad (6)$$

where $f_{ts(n-1)}$ is the tangential spring force at the previous time step, and $\Delta f_{ts}$ is the increment of the tangential force and is given by:

$$\Delta f_{ts} = -k_t \delta_t, \qquad (7)$$

where $k_t$ is the tangential stiffness, and $\delta_t$ is the increment of the tangential displacement. While varying values for the tangential stiffness have been used in the literature, in this study it is set as $2/7k_1$ (Walton and Braun, 1986). The tangential damping force is product of tangential dashpot coefficient, $\beta_t$, and the relative tangential velocity, $v_t$, as given by Eq. (8):

$$f_{td} = -\beta_t v_t. \tag{8}$$

The dashpot coefficient $\beta_t$ is given by:

$$\beta_t = \sqrt{\frac{4m^* k_t}{1 + \left(\frac{\pi}{\ln e}\right)^2}} \tag{9}$$

The limiting tangential friction force is calculated using the Coulombic friction criterion with an additive term $f_o$ and $k_{adh}$, so that the observed friction is given by:

$$|f_{ct}| \leq \mu |f_{hys} + k_{adh}\delta^n - f_o| \tag{10}^*$$

where $f_{ct}$ is the limiting tangential force, $f_n$ is the normal contact spring force and $\mu$ is the friction coefficient. For the torque calculation, the default EDEM rolling friction model is adopted in this study. The total applied torque, $\tau_i$, is given by:

---

[*] Different DEM codes have treated the limiting sliding friction differently, either as a function of only the spring force as used in this paper, or a function of the total force including the damping force. There is not an issue for quasi-static situations as in this study and the readers should consider carefully when it comes to more dynamic situations.

$$\tau_i = -\mu_r \mid f_{hys} \mid R_i \, \omega_i, \tag{11}$$

where $\mu_r$ is the coefficient of rolling friction, $R_i$ is the distance from the contact point to the particle centre of mass and $\omega_i$ is the unit angular velocity of the object at the contact point.

The DEM model was used to simulate uniaxial confined and unconfined loading of cohesionless and cohesive particulate solid. The next section describes scaling of cohesionless and cohesive systems.

*2.2 Scaling of cohesionless system*

To maintain the mechanical and dynamic similarity, the contact model should be scale invariant. However, in linear spring contact model in 3D, the force displacement relationship is dependent on the size of the particle and is not scale invariant (Feng et al., 2007). Therefore contact stiffness needs to be scaled with radius of the particle. For the oblique impact of elastic spheres, Maw et al., (1976) provided a solution that relates contact normal stiffness to the radius of the particles as:

$$k_n = \frac{16}{15} R^{*\frac{1}{2}} E^* \left( \frac{15 m^* V^2}{16 E^* R^{*\frac{1}{2}}} \right)^{\frac{1}{5}} \tag{12}$$

where $m^*$ is the equivalent mass, $R^*$ is the equivalent radius, $E^*$ is the equivalent Young's modulus, and V is a typical impact velocity. In the equation above if mass is expressed in terms of radius, then the contact normal stiffness becomes

$$k_n = C R^* \rho^{*1/5} E^{*4/5} \tag{13}$$

where *C* is a constant. The equation 13 suggests contact normal stiffness should scale linearly with particle radius. In another study, Potyondy and Cundall (2004) assumed a linear relationship between particle size and Young's modulus as:

$$k_n = 2E^* R^* \qquad (14)$$

The commercial code EDEM developed by DEM Solutions uses solution provided by Maw et al. (1976) and another commercial code PFC3D developed by Itasca uses the solution provided by Potyondy and Cundall (2004):

$$k_n = \frac{\pi R_{avg} E}{2} \qquad (15)$$

where $R_{avg}$ is the average radius of contacting particles and E is the Young's modulus. Regardless of different scaling relationship purposed by researchers, it is clear that normal contact stiffness scales linearly with radius of the particle in linear spring contact model. However, the stiffness in Hertz-Mindlin contact model is scale invariant for 3D (Feng et al., 2007). No literature can be found for scaling of unloading stiffness for the case of elasto-plastic contact model.

The scaling relationships for other DEM parameters including tangential stiffness, damping constant, density, sliding friction, and rolling friction is discussed herein.

The tangential contact stiffness $k_t$ has not received the same attention as the normal stiffness $k_n$. Mindlin and Deresiewicz (1953) has proposed that the ratio of normal to tangential stiffness is a material property and independent of size of the particles. This would require tangential stiffness to scale linearly with the radius of the particle. However M&D's study was based on elastic contact of frictional sphere, it is not so clear how $k_t$ will scale with respect to radius of elasti-plastic particles.

The damping constant may have effect in dynamic cases, however, for quasi-static simulation (Midi, 2004) such as ours the effect of damping will not be significant (Obermayr et al., 2011). Moreover, it was proved by Kruggel-Emden et al. (2010) that while compressing the sample at relatively lower rate, dynamic effects are of a smaller importance. In dynamic cases damping constant can be scaled linearly with radius of the particle as suggested by Bierwisch et al., (2009). For the scaled system to reproduce same mechanical behaviour, the density of gravitational potential energy should be same as in the original system. The density of gravitational potential energy is independent of particle radius if porosity of the system is constant. This requires particle density in the original system and scaled system to be the same. Sliding friction is invariant with respect to scaling (Pöschel et al., 2001). Rotational motion (rolling friction) is not scaled in this study.

*2.3 Scaling of cohesive system*

Van der Waals forces are a primary source of adhesion in fine sized particles. In DEM modeling of fine particles, the van der Waals attractive forces are commonly represented using the theoretical adhesive elastic force models such as the JKR and the DMT models (Derjaguin et al., 1975; Johnson et al., 1971) which relates the van der Walls attraction to the radius of the particle as:

$$f_{0DMT} = -4\pi\gamma R^* \quad (16)$$

$$f_{0JKR} = -3\pi\gamma R^* \quad (17)$$

where $\gamma$=surface energy per unit contact area (J/m$^2$). For plastic contacts, Thornton and Ning (1998) suggested that plastic deformation at the contact causes an increase in pull-off force approximately by a factor of 2 compared to the elastic JKR model with the pull-off force given by:

$$f_{0plastic} \approx -6\pi\gamma R^* \quad (18)$$

According to Israelachvili (1992), the pull-off force between two approaching spheres of equal diameter is given by:

$$f_0 = \frac{AR^*}{6s^2} \quad (19)$$

where $A$=Hamaker's constant and $s$=separation between the particles. Hamaker's constant (A) can be expressed as:

$$A = \pi^2 C_f \rho_a^2 \quad (20)$$

where $C_f$ is a constant and $\rho_a$ is number of atoms per unit volume of contacting bodies and is a material property. The above deductions suggest that the adhesive force is linearly proportional to the radius of the particle.

This is in contrast to the proposition of Rumpf (1962) where the relationship between tensile strength ($\sigma_t$), and the inter-particle contact force ($f_0$) for a system of hard mono-disperse sphere with a random isotropic packing is given by the following equation:

$$f_0 = \frac{4\pi R^2}{\phi k} \sigma_t \quad (21)$$

in which $\phi$ = packing fraction, and $k$ = co-ordination number.

This suggests that the inter-particle contact force should scale with the square of particle radius. In addition, as particle radius decreases, contact surface area per unit volume of particle increases. Since adhesive forces are related to the surface area of a particle and since the surface area is proportional to the square of the radius of the particle, this suggests that adhesion force is quadratically proportional to the radius of particle.

Some researchers suggest keeping the bond number ($f_0/f_g$) the same in original and scaled system, where $f_g$ is the gravitational force that is equal to the weight of the particle and is expressed as:

$$f_g = \frac{4}{3}\pi R^3 \rho \quad (22)$$

where $\rho$ = density of the particle. Therefore, $\frac{f_0}{f_g} \propto R^3$. This suggests $f_0$ should be scaled cubically with the radius of the particle. The different approaches outlined in Equations 16-22 suggest that adhesive force may scale linearly, quadratically, and cubically with particle radius. These are investigated for a cohesive system below.

## 3. Simulation set-up

The computer simulations reported here consider a series of uniaxial compression tests in a rectangular cuboid of 50 mm thickness (>6*diameter of largest particle), 150 mm width, and 300 mm height (see Figure 2). Periodic boundaries were used along X and Y direction to avoid the wall boundary effect. The cuboid contains a top and a bottom plate. A series of uniaxial compression simulations were conducted using the simplified DEM contact model. Each simulation consisted of several stages of loading: a) filling the cuboid; b) confined consolidation to a 40kPa vertical stress level and subsequent unloading with periodic boundary, c) and finally unconfined compression of the sample to failure without periodic boundary after the removal of the confining mould. The random rainfall method was adopted to provide a random packing of particles. For cohesionless case, similar porosities were achieved for different size particles by vibrating the system with frequency of 60 Hz and amplitude of 1.5mm for simulation time of 2s. Loading only commenced when the system has reached a quasi-static state, as indicated by the kinetic to potential energy ratio at less than $10^{-5}$ with a constant coordination number. For cohesive system, it was difficult to get reproducible porosity in fill stage; therefore the porosity corresponding to an initial vertical stress of 5 kPa was considered as initial packing for subsequent loading.

Compression was achieved by moving the top plate at a constant speed until a desired bulk vertical stress was attained. Subsequently, unloading was performed by an upward retreat of the upper plate. The confining periodic boundaries were then removed and the unconfined samples were allowed to reach the new equilibrium, and finally the top platen was lowered to fail the sample. The loading and unloading were performed at an axial speed of 10 mm/s (strain rate<$0.1s^{-1}$) throughout to ensure quasi-static loading. The quasi-static loading was confirmed by inertia number being less than $1\times10^{-4}$ (Midi, 2004) in all simulations. The lower plate remained stationary in all stages.

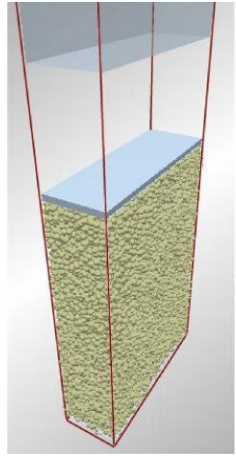

**Figure 2 Simulation set-up**

Three special cases of the linear contact model (Figure 1b) proposed as shown in Figure 3 are explored here. The scaling law was first applied for the cohesionless case (case I), and for the constant adhesion case (case II), and finally for the load dependent adhesion case (case III).

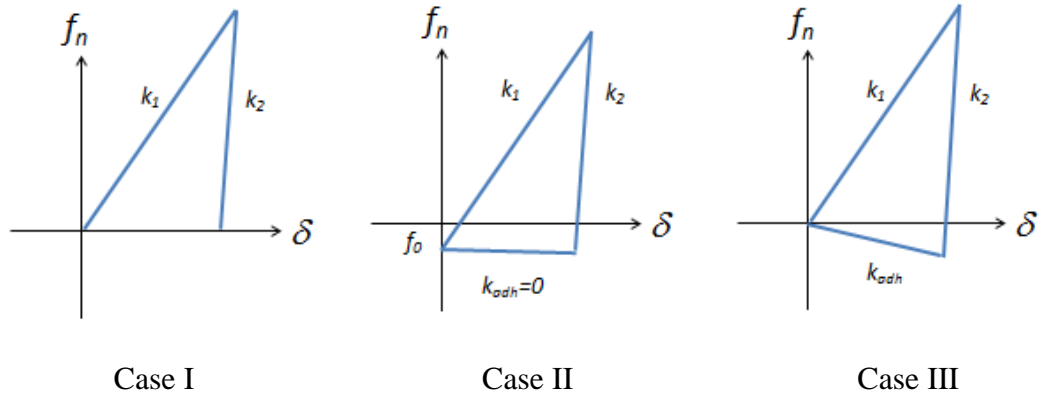

Figure 3 Different cases of simulated contact model.

For simplicity, the particle shape used in this study was spherical and of uniform size in each simulation. The cohesive contact model was only applied to particle-particle interactions. The particle-geometry interactions were modelled using the Hertz-Mindlin (no-slip) contact modelwith no particle-geometry adhesion.

**Table 1. Input parameters**

| Parameter | Value |
|---|---|
| Particle density, $\rho$ (kg/m$^3$) | 2000 |
| Loading spring stiffness, $K_1$ (N/m) | $5 \times 10^3$ to $1 \times 10^4$ |
| Unloading spring stiffness, $K_2$ (N/m) | $2.5 \times 10^4$ to $5 \times 10^4$ |
| Load dependent stiffness, $K_{adh}$ (N/m) | $5 \times 10^3$ to $7.5 \times 10^3$ |
| Adhesion force, $f_0$ (N) | 0 to -1.6 |
| Tangential stiffness, $K_t$ (N/m) | 2/7 $K_1$ |
| Particle static friction, $\mu_{sf}$ | 0.5 |
| Particle rolling friction, $\mu_{rf}$ | 0.001 |
| Particle radius (R), mm | 2.5 to 3.75 |
| Top and bottom platen friction, $\mu_{Pf}$ | 0.3 |
| Simulation time step (s) | $1 \times 10^{-5}$ |

## 4. Simulation results

### *4.1 Cohesionless system*

The axial stressstrain response and the corresponding stress-porosity behaviour during the confined loading and unloading simulation are shown in Figure 4 and Figure 5 respectively. The simulation with particle radius of 2.5 mm is taken as the reference case and this was compared with the case with particle radius of 3.75mm scaled and unscaled. The particle density and sample porosity were kept the same throughout to keep the gravitational potential energy the same in both the large particle and the small particle systems. For the unscaled (all model parameters unchanged) 3.75mm case, it can be clearly seen that increasing the particle size without scaling the stiffness produces a softer bulk response compared to the reference case. However, when stiffness was scaled linearly with the particle radius, the stress-strain response and the corresponding porosity-stress response for the 3.75mm particle converged to that for the reference case of 2.5mm particle.

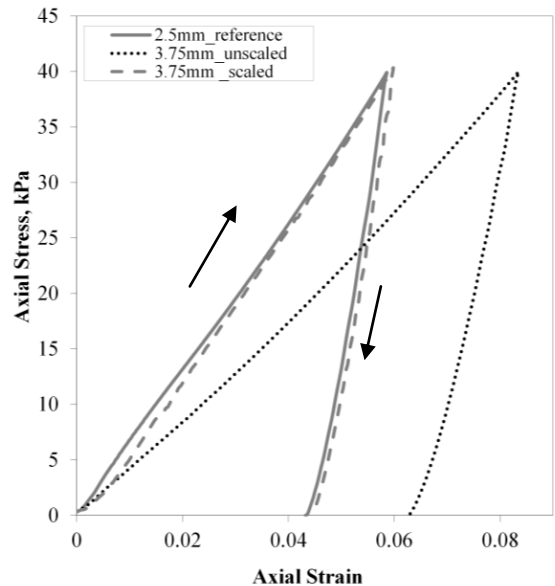

**Figure 4 Confined compression: Axial strain vs axial stress**

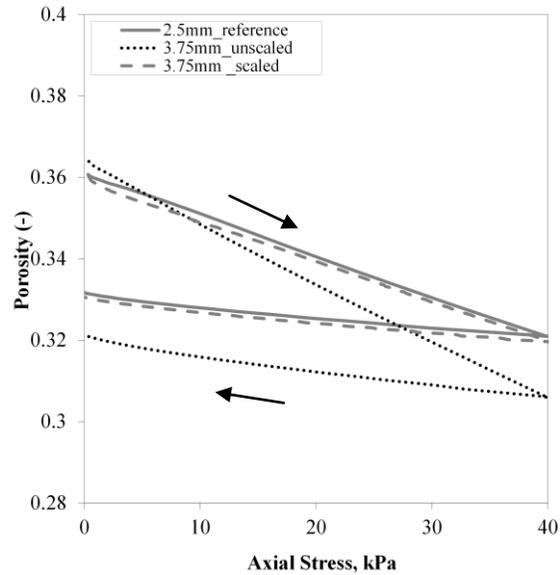

**Figure 5 Confined compression: Axial stress vs. porosity**

The variation of porosity across the sample height was also investigated in Figure 6. It can be seen that the porosity was very similar for the reference case and the scaled case, however, without scaling of the contact stiffness, the porosity was consistently smaller throughout the height of the sample. It thus follows that increasing the particle size without scaling the stiffnesses will result in a softer system.

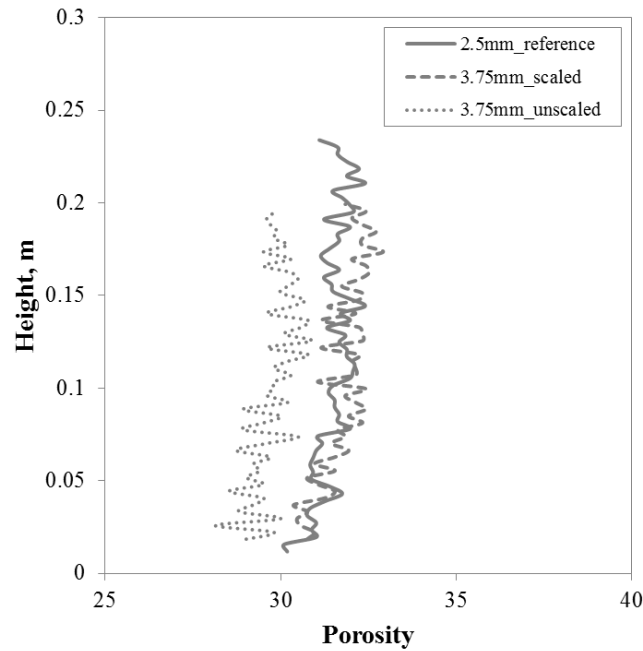

**Figure 6 Porosity variation plotted against the height**

The effect of scaling stiffness was also investigated on microstructural coordination number. The coordination number (CN) for each of the three simulations are compared in Figure 7. The results show that the CN during the loading and unloading also evolved in the same fashion to the reference case when the stiffness scaling was deployed. For the unscaled case, the CN increased at a higher rate compared to the reference case, showing a significant change in the internal contact fabric of the assembly.

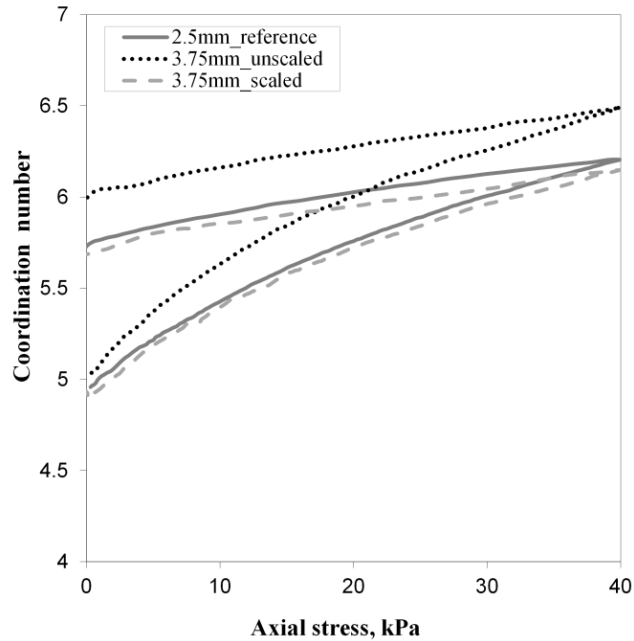

**Figure 7 Evolution of coordination number**

It can therefore be concluded that the loading and unloading bulk response for a cohesionless particle assmbly can be particle scale invariant by a linear scaling of the contact normal and tangential stiffnesses with the particle radius.

*4.2 Cohesive system*

### 4.2.1 Constant adhesion

For the cohesive system, the normal and tangential stiffness (both loading and unloading) were scaled linearly as in the cohesionless system. Additionally, linear, quadratic, and cubic scaling of the adhesive force parameter $f_o$ with particle radius was explored. The simulation with particle size of 2.5 mm was again the reference case. The axial stressstrain response and the corresponding porosity-stress response are shown in Figure 8 and Figure 9 respectively for different particle sizes with different scaling approaches for the adhesive force. The Figure 9 shows sthat when the adhesive force was scaled linearly with particle size, the initial porosity at 5kPa stress level was found to be lower when compared to the quadratic and cubic scaling. The linear scaling thus produced a denser initial packing resulting in less compression under loading than the quadratic and cubic scaling as shown in the stress-strain curve (Figure 8). Conversely the cubic scaling of adhesive force with particle size produced a higher initial porosity and the sample compressed the most during loading. The quadratic scaling of adhesive force with particle size produced very similar stress-porosity and stress-strain response for particle size in a range of 2 to 3.75 mm.

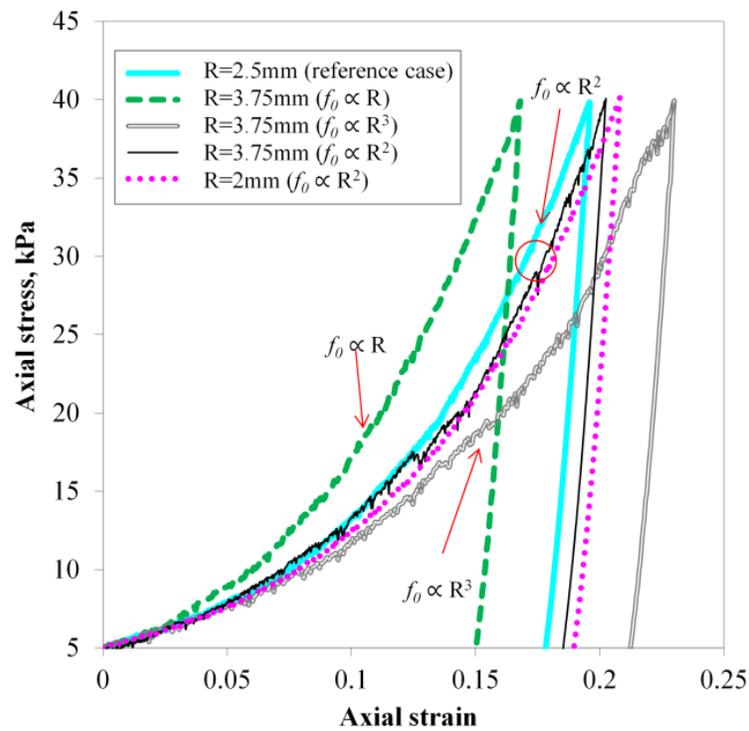

**Figure 8 Confined compression: Axial stress vs strain**

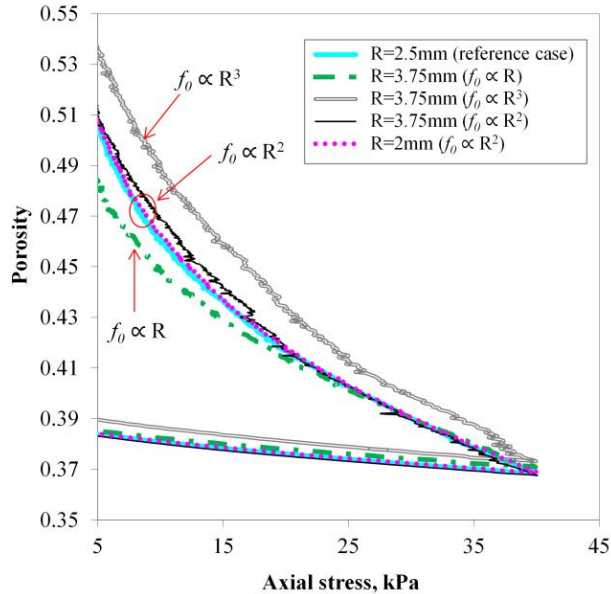

**Figure 9 Confined compression: Porosity vs axial stress**

The scaling of the adhesive force was further examined by looking into the unconfined compression behavior to failure as shown in Figure 10. The quadratic scaling produced very similar unconfined stress-strain (initial stiffness and unconfined strength) behaviour to shear failure for different sized particles of 2-3.75mm. The linear scaling with particle size underestimated the unconfined strength and the cubic scaling overestimated the strength. This confirms that scaling the cohesive force by keeping the bond number constant (i.e cubic scaling) is not the right strategy.

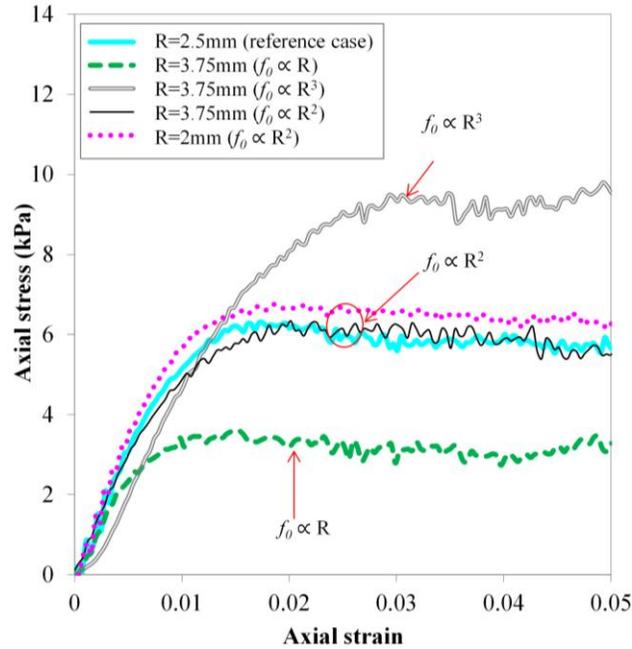

**Figure 10** Unconfined compression: Axial stress vs strain

The above analysis has clearly shown that adhesive force scales quadratically with the particle radius. This is consistent with results from Walton and Johnson (2009a) on the DEM simulations of rotary drum flows using their previously implemented DEM code (Walton and Johnson, 2009). They found that the scaling of the pull-off force with the square of the particle size produced flows that were qualitatively in agreement. Bierwisch et al. (2009) also found in simulations of rapid granular flow from a moving container and angle of repose formation that adhesive force scales with square of the radius of the particles. According to our study the combined linear scaling of the spring contact stiffness and quadratic scaling of the adhesive force parameter appear to be a robust strategy for the upscaling of particle size.

Figure 11shows reduction in simulation time with decreasing size of particles. More than seven fold decrease in computational time was observed if particle size is scaled from 2 mm to 3.75 mm for the simulation of uniaxial compression using 12 core processors in this study. With the increase in particle size, the stiffness is increased however the number of particles decreases significantly and causes significant reduction in computational time. The scaling laws allow the use of larger particle sizes whilst reproducing similar mechanical response of a particulate assembly with smaller particles and help to reduce the computational time significantly. It is important to note that although particle size in a small range of 2-3.75 mm were investigated, in terms of number of particles (n) there is almost an order of magnitude difference (5100 vs 34500). It is expected that the scaling should be valid for smaller sizes.

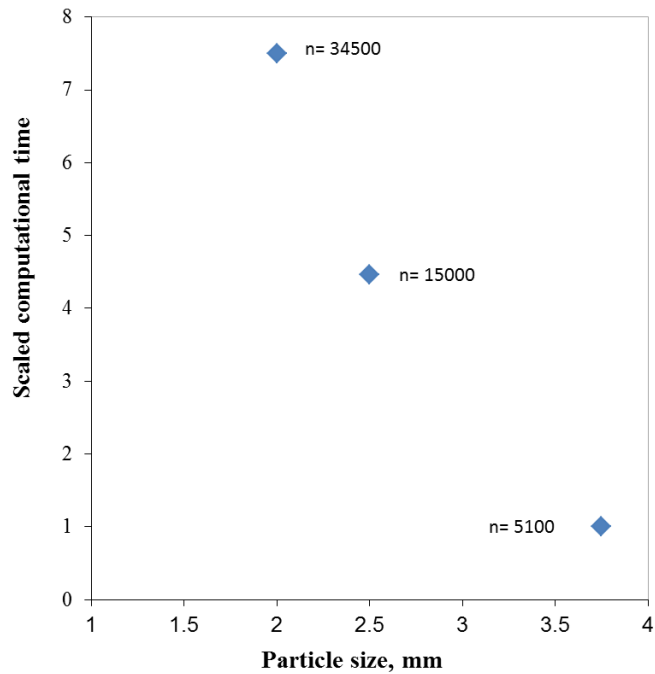

**Figure 11 Reduction in computational time with scaling**

**4.2.2 Load dependent adhesion**

In this section the scaling of load dependent stiffness ($k_{adh}$) with zero $f_o$ (case III- Figure 3) is explored. The normal loading and unloading stiffness and tangential stiffness are scaled linearly as established in previous section. Additionally, the $k_{adh}$ is scaled linearly with the radius of the particle. Figure 12 and Figure 13 show stress strain and corresponding porosity stress behaviour during confined compression, respectively. Similar stress strain response with small discrepancy in peak strain can be observed for scaled and reference case. The slightly lower peak strain for the scaled case can be attributed to slightly lower initial porosity arising from random generation of particles. Although a small difference in initial porosity for scaled and reference sample can be seen, both curves converge at higher stress (Figure 13).

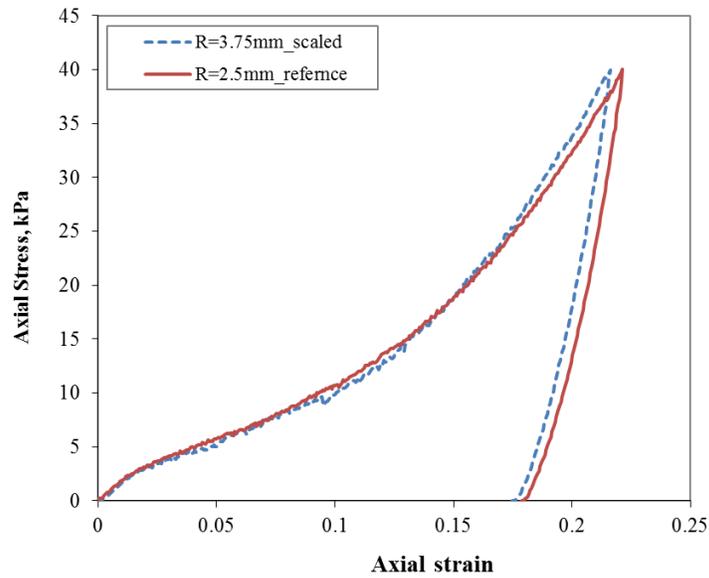

**Figure 12 Confined compression: Axial stress vs strain**

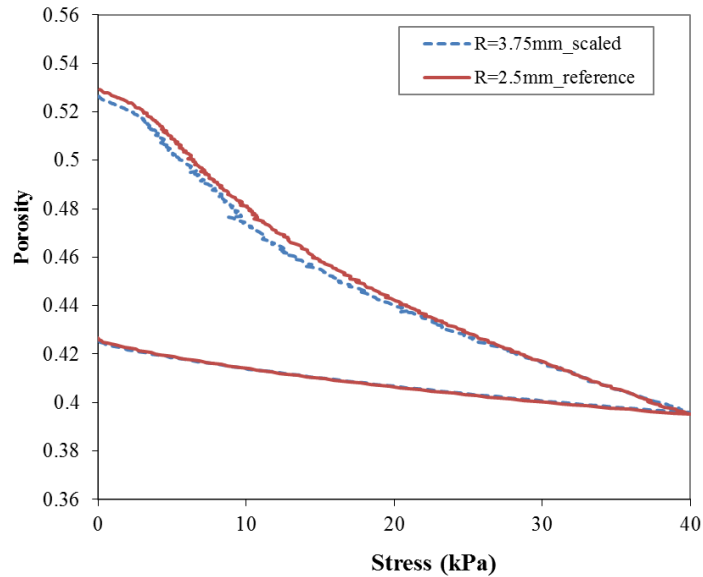

**Figure 13 Confined compression: Porosity vs axial stress**

Figure 14 shows stress strain response during unconfined compression. The initial stiffness during unconfined compression are almost identical for the both cases, however, the maximum strength for scaled case was 9.5% lower than that for reference case. The low strength associated with scaled case was found to be related with lower CN. After the end of consolidation when confinement is removed, CN drops. The drop in CN was higher for the scaled case, although the CN at the end of consolidation was the same in both cases.

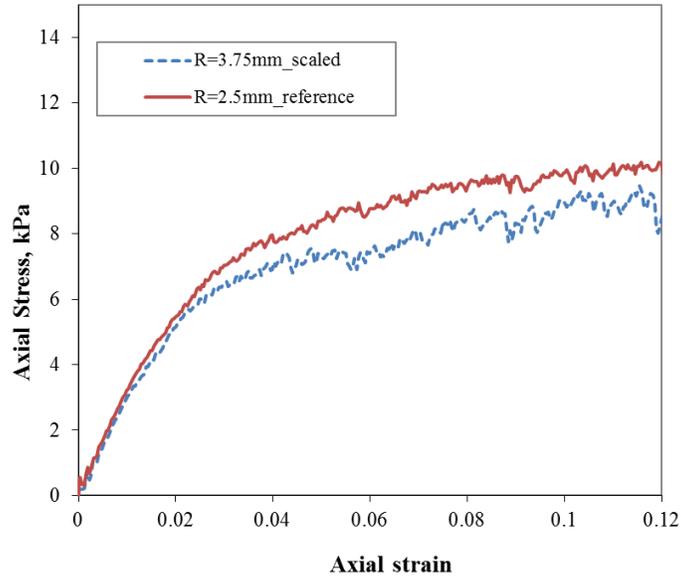

Figure 14 Unconfined compression: Axial stress vs axial strain

## 5. Conclusion

A study of the scaling laws to produce scale independent computations of confined compression and unconfined loading has been presented. In the linear spring model with elasto-plastic deformation and no cohesion, the contact loading and unloading stiffness (normal and tangential) scales linearly with particle size. A very good agreement in the macroscopic (stress-strain and stress-porosity relations) and the microscopic (stress-coordination number relation) behaviour was found for different particle sizes when the contact stiffness was scaled linearly. For simulation with a constant adhesion, the scaling of the adhesion force parameter with the square of the particle radius (2~3.75mm in this study) produced confined stress-strain and stress-porosity behaviour, and unconfined stress-strain behaviour that remained remarkably similar as the size of the particles were increased. Furthermore, linear scaling of load dependent stiffness with the radius of particle produced very similar confined stress-strain and corresponding stress-porosity relation. Also almost identical stiffness during unconfined compression was found. However, the unconfined strength for scaled system was within 10% of that for the reference system. Thus, by scaling the stiffness linearly and adhesive force quadratically, a DEM model using larger particle size can exhibit the same bulk properties as the system with small particle size. This scaling may have limitations when length scale of the particle size becomes comparable to the length scale of system. Nevertheless, such scaling laws are particularly useful for studying very large scale particulate systems where scaling up the particle size will result in considerably less computational effort.

## 6. Acknowledgments

The support from the EU Marie Curie Initial Training Network is gratefully acknowledged. The authors would also like to thank J-F Chen, Carlos Labra, and J.P. Morrissey for many useful discussions.

## 7. References

Bierwisch, C., Kraft, T., Riedel, H., Moseler, M., 2009. Three-dimensional discrete element models for the granular statics and dynamics of powders in cavity filling. J. Mech. Phys. Solids 57, 10–31. doi:10.1016/j.jmps.2008.10.006

Chaudhuri, B., Mehrotra, A., Muzzio, F.J., Tomassone, M.S., 2006. Cohesive effects in powder mixing in a tumbling blender. Powder Technol. 165, 105–114. doi:10.1016/j.powtec.2006.04.001

Cundall, P.A., Strack, D.L., 1979. A discrete numerical model for granular assemblies. Geotechnique 1, 47–65.

DEM Solutions Ltd., 2010. EDEM 2.3 Programming Guide, Revision 3. ed, Online. Edinburgh, Scotland, UK.

Derjaguin, B., Muller, V.M., Toporov, Y.P., 1975. Effect of contact deformations on the adhesion of particles. J. Colloid Interface Sci. 53, 314–326.

Ebrahimi, M., Crapper, M., Ooi, J.Y., 2014. Experimental and Simulation Studies of Dilute Horizontal Pneumatic Conveying. Part. Sci. Technol. 32, 206–213. doi:10.1080/02726351.2013.851133

Feng, Y., Han, K., Owen, D., Loughran, J., 2007. Upscaling of discrete element models for particle systems, in: Proceedings of the 4 Th Int. Conf. on Discrete Element Methods. pp. 27–29.

González-Montellano, C., Ayuga, F., Ooi, J.Y., 2011. Discrete element modelling of grain flow in a planar silo: influence of simulation parameters. Granul. Matter 13, 149–158. doi:10.1007/s10035-010-0204-9

Israelachvili, J.N., 1992. Intermolecular and surface forces. San Diego Acad.

Johnson, K.L., Kendall, K., Roberts, A.D., 1971. Surface energy and the contact of elastic solids. Proc. R. Soc. London. A. Math. Phys. Sci. 324, 301.

Jones, R., 2003. From Single Particle AFM Studies of Adhesion and Friction to Bulk Flow: Forging the Links. Granul. Matter 4, 191–204.

Ketterhagen, W.R., Curtis, J.S., Wassgren, C.R., Hancock, B.C., 2009. Predicting the flow mode from hoppers using the discrete element method. Powder Technol. 195, 1–10. doi:10.1016/j.powtec.2009.05.002

Kruggel-Emden, H., Stepanek, F., Munjiza, A., 2010. A study on adjusted contact force laws for accelerated large scale discrete element simulations. Particuology 8, 161–175. doi:10.1016/j.partic.2009.07.006

Luding, S., 2008. Cohesive, frictional powders: contact models for tension. Granul. Matter 10, 235–246.

Maw, N., Barber, J., Fawcett, J., 1976. The oblique impact of elastic spheres. Wear.

Midi, G.D.R., 2004. On dense granular flows. Eur. Phys. J. E. Soft Matter 14, 341–65. doi:10.1140/epje/i2003-10153-0

Mindlin, R.D., Deresiewicz, H., 1953. Elastic spheres in contact under varying oblique forces. J. Appl. Mech 20.

Mio, H., Akashi, M., Shimosaka, A., Shirakawa, Y., Hidaka, J., Matsuzaki, S., 2009. Speed-up of computing time for numerical analysis of particle charging process by using discrete element method. Chem. Eng. Sci. 64, 1019–1026. doi:10.1016/j.ces.2008.10.064

Moreno-atanasio, R., Antony, S.J., Ghadiri, M., 2005. Analysis of flowability of cohesive powders using Distinct Element Method. Powder Technol. 158, 51–57.

Obermayr, M., Dressler, K., Vrettos, C., Eberhard, P., 2011. Prediction of draft forces in cohesionless soil with the Discrete Element Method. J. Terramechanics 48, 347–358. doi:10.1016/j.jterra.2011.08.003

Parteli, E.J.R., Schmidt, J., Blumel, C., Wirth, K.-E., Peukert, W., Poschel, T., 2014. Attractive particle interaction forces and packing density of fine glass powders. Sci. Rep. 4.

Pöschel, T., Salueña, C., Schwager, T., 2001. Scaling properties of granular materials. Phys. Rev. E 64, 1–4. doi:10.1103/PhysRevE.64.011308

Potyondy, D.O., Cundall, P. a., 2004. A bonded-particle model for rock. Int. J. Rock Mech. Min. Sci. 41, 1329–1364. doi:10.1016/j.ijrmms.2004.09.011

Rumpf, H., 1962. The strength of granules and agglomerate, in: Knepper, W. (Ed.), Agglomeration. Wiley Interscience, New York.

Sakai, M., Koshizuka, S., 2009. Large-scale discrete element modeling in pneumatic conveying. Chem. Eng. Sci. 64, 533–539. doi:10.1016/j.ces.2008.10.003

Sheng, Y., Lawrence, C.J., Briscoe, B.J., Thornton, C., 2004a. Numerical studies of uniaxial powder compaction process by 3D DEM. Eng. Comput. 21, 304–317. doi:10.1108/02644400410519802

Sheng, Y., Lawrence, C.J., Briscoe, B.J., Thornton, C., 2004b. Numerical studies of uniaxial powder compaction process by 3D DEM. Eng. Comput. 21, 304–317.

Singh, A., Magnanimo, V., Saitoh, K., Luding, S., 2014. Effect of cohesion on shear banding in quasistatic granular materials. Phys. Rev. E 90, 22202.

Singh, A, Magnanimo, V., and Luding, S., 2015. A contact model for sticking of adhesive mesoscopic particles. Under Rev. Powder Technol. 74.

Thakur, S.C., Ahmadian, H., Sun, J., Ooi, J.Y., 2014. An experimental and numerical study of packing, compression, and caking behaviour of detergent powders. Particuology 12, 2–12.

Thakur, S.C., Morrissey, J.P., Sun, J., Chen, J.F., Ooi, J.Y., 2014. Micromechanical analysis of cohesive granular materials using the discrete element method with an adhesive elasto-plastic contact model. Granul. Matter 16, 383–400. doi:10.1007/s10035-014-0506-4

Thornton, C., Ning, Z., 1998. A theoretical model for the stick/bounce behaviour of adhesive, elastic-plastic spheres. Powder Technol. 99, 154–162.

Walton, O.R., Braun, R.L., 1986. Viscosity, granular-temperature, and stress calculations for shearing assemblies of inelastic, frictional disks. J. Rheol. (N. Y. N. Y). 30, 32.

Walton, O.R., Johnson, S.M., 2009. Simulating the effects of interparticle cohesion in micron-scale powders, in: AIP Conference Proceedings. Golden, Colorado, pp. 897–900. doi:10.1063/1.3180075

Walton, O.R., Johnson, S.M., 2010. DEM Simulations of the effects of particleshape, interparticle cohesion, and gravity on rotating drum flows of lunar regolith, in: Earth and Space. Honolulu, Hawaii, pp. 1–6.

Xu, B., 1997. Numerical simulation of the gas-solid flow in a fluidized bed by combining discrete particle method with computational fluid dynamics. Chem. Eng. Sci. 52, 2785–2809. doi:10.1016/S0009-2509(97)00081-X

Yen, K., Chaki, T., 1992. A dynamic simulation of particle rearrangement in powder packings with realistic interactions. J. Appl. Phys. 71, 3164–3173.

Zhu, H., Zhou, Z., Yang, R., Yu, A., 2008. Discrete particle simulation of particulate systems: A review of major applications and findings. Chem. Eng. Sci. 63, 5728–5770. doi:10.1016/j.ces.2008.08.006